\documentclass[12pt,a4paper]{article}

\usepackage{latexsym}
\usepackage{amssymb}
\usepackage{amsmath}

\newfont{\bl}{msbm10 scaled \magstep1}
\newfont{\bs}{msbm8}
\newfont{\bss}{msbm7}

\numberwithin{equation}{section}

\newcommand{\R}{\mbox{{\bl R}}}

\newcommand{\ZZ}{\mbox{{\bs Z}}}

\newcommand{\SI}{\mbox{{\bl S}}}
\newcommand{\eps}{\varepsilon}

\newcommand{\cD}{{\cal D}}
\newcommand{\cE}{{\cal E}}

\newcommand{\cG}{{\cal G}}

\newcommand{\cN}{{\cal N}}
\newcommand{\cO}{{\cal O}}

\newcommand{\cV}{{\cal V}}
\newcommand{\cW}{{\cal W}}
\newcommand{\xx}{x_{12}}

\begin{document}
\vspace{1cm}
\begin{center}

{\bf \LARGE Generalized functions for quantum
        \vspace{1.5mm}\\ 
         fields obeying quadratic exchange
        \vspace{3mm}\\
         relations}
        \vspace{8mm}\\
        {\large H. Grosse}$\,$\footnote{Part of project P 11783-PHY of the
          ``Fonds zur F\"orderung der wissenschaftlichen Forschung in
          \"Osterreich''}
          \vspace{1mm}\\
	  {\small Institut f\"ur Theoretische Physik,
          Universit\"at Wien}\vspace{1mm}\\
          {\small A-1090 Vienna, Austria}
          \vspace{4mm}\\
        {\large M. Oberguggenberger}\vspace{1mm}\\
	  {\small Institut f\"ur Mathematik und Geometrie,
	  Universit\"at Innsbruck}\vspace{1mm}\\
          {\small A-6020 Innsbruck, Austria}
          \vspace{4mm}\\
        {\large I. T. Todorov}$\,$\footnote{Permanent address:
          Institute for Nuclear Research and Nuclear Energy,
          Bulgarian Academy of Sciences, BG-1784 Sofia, Bulgaria}
          \vspace{1mm}\\
       	  {\small Erwin Schr\"odinger International Institute for Mathematical Physics}
          \vspace{1mm}\\
          {\small A-1090 Vienna, Austria}
\end{center}
\vspace{5mm}

\begin{abstract}
The axiomatic formulation of quantum field theory (QFT) of the 1950's in
terms of fields defined as operator valued Schwartz distributions is
re-examined in the light of subsequent developments. These include, on the
physical side, the construction of a wealth of (2-dimensional) soluble
QFT models with quadratic exchange relations, and, on the mathematical
side, the introduction of the Colombeau algebras of generalized functions.
Exploiting the fact that energy positivity gives rise to a natural 
regularization of Wightman distributions as analytic functions in a tube
domain, we argue that the flexible notions of Colombeau theory which can 
exploit particular regularizations is better suited (than Schwartz
distributions) for a mathematical formulation of QFT.
\end{abstract}

\section{Introduction}

The only mathematically precise notion of a quantum field available to date,
a field as an operator valued distribution, was coined (by Wightman,
Bogolubov and others) in the 1950's (for reviews and references to original
work, see \cite{Stre, Bog1}). It made use of the then new and fashionable 
Schwartz distribution theory \cite{Sch2}. The concept proved indeed natural
and useful for problems in which linear functional analysis is applicable,
like in the theory of free fields. For instance, the canonical equal time
commutation relations between a (scalar) field $\varphi(t,{\bf x})$ and
its conjugate momentum $\pi(t,{\bf y})$ is written in terms of the
3-dimensional Dirac $\delta$-function:
\begin{equation}
   [\varphi(t,{\bf x}), \pi(t,{\bf y})] = i\,\delta({\bf x} - {\bf y})
   \Longleftrightarrow 
   [\varphi(t,f), \pi(t,g)] = i\int\! f({\bf x})g({\bf x})\,d{\bf x}\,,
\end{equation}
where $\varphi(t,f)$ is the smeared (sharp time) field: 
\begin{equation}
   \varphi(t,f) = \int\! f({\bf x})\varphi(t,{\bf x})\,d{\bf x}\ \ 
                  (d{\bf x} = d^3x) 
\end{equation}
with $f$ belonging to a suitable class of (smooth, falling at infinity) 
test functions. Difficulties appeared as soon as an interacting theory was
considered in which products of distributions with coinciding arguments
were encountered. Nevertheless, perturbative renormalization was successfully
treated in the same conceptual framework, using continuation of linear
functionals (originally defined on a subspace of test functions which
vanish for coinciding arguments - for a review, see \cite{Bog2} and
\cite{Coll}).\\

In the four decades that elapsed after the first achievements of axiomatic 
QFT, new types of (soluble) QFT models (in particular,
in 2-dimensional conformal field theory \cite{Fran}) and a new notion of
generalized functions \cite{Colo,Rosi} have emerged. The aim of this note is
to re-examine the role of generalized functions in QFT in the light of
these developments.\\

Two (and one) dimensional integrable models typically involve {\em quadratic
exchange relations} instead of the canonical commutation relations (1.1).
For fields with non-(half-)integer dimensions the corresponding $R$-matrix
(the phase factor in the simplest ``anyonic'' Thirring model \cite{Thir,Ilie})
depends on the sign of the difference $x-y$ between two points (more
precisely, it depends on the path along which $x$ and $y$ are interchanged).
Combined with the equations of motion this discontinuous phase factor yields
inconsistencies if viewed as a Schwartz distribution. We demonstrate that it
can be given a (consistent) unambiguous meaning if considered instead as a
Colombeau generalized function.\\

The Colombeau theory is adapted to describing singular functions with a
preferred regularization. This is the case with the Wightman functions
\begin{equation}
     \cW (x_1-x_2, \dots , x_{n-1}-x_n) 
      = \langle 0|\phi(x_1)\dots\phi(x_n)|0\rangle
\end{equation}
in any QFT satisfying the {\em spectrum condition} (i.e. the basic
assumption of {\em energy positivity} of physical states). Under this
assumption the distributions (1.3) can be viewed as boundary values of
analytic functions $\cW(\zeta_1,\dots, \zeta_{n-1})$ holomorphic in a
tube domain. For a chiral field in a 2-dimensional (conformally invariant)
QFT, the chief example of the present paper, $x_j - x_{j+1}$ are
1-dimensional light-like variables $(x_j = x^1_j - x^0_j)$ and
$\cW(\zeta_1,\dots, \zeta_{n-1})$ is analytic for $\Im \zeta_j > 0$.
This analytic function provides the preferred regularization for the 
vacuum expectation values of fields' products. In particular, for a
2-point function, the (scale) invariant Wightman distribution
$(0 - ix)^{-\lambda}$ is associated with the {\em Colombeau sequence}
\begin{equation}
   (\eps - ix)^{-\lambda},\; \eps \searrow 0\ (\lambda > 0)\;.
\end{equation}

The paper is organized as follows. In Sec. 2 we review the basic 
ingredients of the chiral Thirring model. The charged fields satisfy
anyonic exchange relations whose singularity structure is the same
as that of more general $R$-matrix relations in a non-abelian current
algebra model. In Sec. 3 we exhibit the difficulties of giving a precise 
meaning to these relations in terms of distributions and demonstrate 
how these difficulties are overcome in the Colombeau framework.
An informal (physicist oriented) synopsis of the Colombeau theory
is given in an Appendix.
%
\section{The chiral Thirring model: a synopsis}
%
The chiral Thirring model \cite{Thir,Ilie} can be viewed as the theory
of a (massless) charged field $\psi(x,g)$ and its conjugate
$\psi(x,g)^\ast = \psi(x,-g)$ coupled to an $U(1)$-current $j(x)$. Here
$x$ is a lightcone variable and $j$ is the corresponding current
component:
\begin{equation}
  x = x^1 - x^0,\ j = \frac{1}{\sqrt{2}}\left(j^0 + j^1\right)\,.
\end{equation}
In the original formulation of the model $\psi$ is viewed as the basic
field and the current is expressed as a ``normal product'' of two
oppositely charged fields, $j \sim \psi(x,g)\psi(x,-g)$. In Haag's
algebraic approach to local quantum physics \cite{Haag}, which we adopt
here, one starts instead with the local current $j$ that generates the
chiral observable algebra; the charged fields are then constructed as 
local intertwiners among the different superselection sectors of the
current algebra theory (see e.g. \cite{Buch}). We shall summarize the
basic assumptions and results of this approach.\\

1) {\em The current is a free field} of dimension 1 with a {\em scale 
invariant 2-point function}
\begin{equation}
   (2\pi)^2\langle 0|j(x_1)j(x_2)|0\rangle = (0 - i\xx)^{-2};\
    \xx = x_1 - x_2\;.
\end{equation}
The definition of the generalized function on the right hand side (a special
case of the limit (1.4)) exploits {\em energy positivity} - as noted in the
Introduction.\\

2) A {\em primary charged field} $\psi(x,g)$ obeys the {\em equation of 
motion}
\begin{equation}
   i\frac{d}{dx}\psi(x,g) = 2\pi g:\!j(x)\psi(x,g)\!:
\end{equation}
where the current field normal product is expressed in terms of the current's
frequency parts:
\begin{equation}
   :\!j(x)\psi(x,g)\!:\; = j_+(x)\psi(x,g) + \psi(x,g)j_-(x)\ \ (j = j_+ + j_-)\;.
\end{equation}

3) The {\em frequency parts} $j_\pm(x)$ are characterized by the {\em vacuum
conditions}
\begin{equation}
   j_-(x)|0\rangle = 0 = \langle 0|j_+(x)
\end{equation}
and the {\em Ward identities}
\begin{eqnarray}
   2\pi \left[ j_-(x_1), \psi(x_2,-g) \right] 
                         =  \frac{g}{0-i\xx}\psi(x_2,-g)\,, \nonumber \\
   2\pi \left[ \psi(x_2,-g), j_+(x_1) \right] 
                         =  \frac{g}{0+i\xx}\psi(x_2,-g)\,.
\end{eqnarray}

{\bf Proposition 2.1.}$\ $ {\em Requirements 1)2)3) allow to compute the
(analytically continued) n-point correlation functions of charged fields
which, in accordance with charge conservation, are given by}
\begin{eqnarray}
  (2\pi)^{n/2}\langle 0|\psi(x_1 + i\eps_1,g_1)\dots 
               \psi(x_n + i\eps_n,g_n)|0\rangle \nonumber \\
             = \prod_{1\leq j<k\leq n} (\eps_{jk} - ix_{jk})^{g_j g_k}\,
               \delta_{g_1 + \dots + g_n,0}
\end{eqnarray}
{\em for} $\eps_{jk} = \eps_j - \eps_k > 0\,, j < k,$ {\em and} $x_{jk}
= x_j - x_k$.\\

The proof uses the so called {\em vertex operator construction} (for a
complete treatment within Haag's algebraic approach and for references to
earlier work - see \cite{Buch}).\\

As a special case we obtain the 2-point function
\begin{eqnarray}
   2\pi\langle 0|\psi(x_1 + i\eps,-g)\psi(x_2,g)|0\rangle
    = (\eps - i\xx)^{-g^2} \nonumber \\
    = 2\pi\langle 0|\psi(x_1 + i\eps,g)\psi(x_2,-g)|0\rangle\;.
\end{eqnarray}

The correlation functions (2.7) give rise to the following {\em anyonic
exchange relations}
\begin{equation}
   \psi(x_2 + i\eps,g_2)\psi(x_1,g_1)|\Phi\rangle
   = U_\eps(\xx,g_1g_2) \psi(x_1 + i\eps,g_1)\psi(x_2,g_2)|\Phi\rangle\;,
\end{equation}
valid for any finite energy state $\Phi$ (and $\eps > 0$).
Here
$$
   U_\eps(x,\lambda) = \left(\frac{\eps + ix}{\eps - ix}\right)^\lambda
     = e^{{\textstyle i\lambda\sigma_\eps(x)}} \eqno{\mbox{(2.10a)}}
$$
where
$$
   \sigma_\eps(x) = i \ln\frac{\eps - ix}{\eps + ix} 
   \longrightarrow \sigma(x):=
   \left\{ \begin{array}{ll}
            \pi, & x > 0\\
            0, & x = 0 \\
            -\pi, & x < 0
            \end{array} \right.
   \ \mbox{as}\ \eps \searrow 0\;. \eqno{\mbox{(2.10b)}}
$$ 

{\em Remark 2.1.} The existence of an analytic continuation in a tube
domain allows to define the product of Wightman distributions with the
{\em same order of arguments} even within Schwartz distribution theory.
The problem arises in giving a precise meaning to the exchange relations 
since then we effectively have to multiply such distributions with an
oppositely ordered pair of components (or stated differently, to consider
the product of a discontinuous function with a Wightman distribution).
It is important that in this more general situation, too, analytic
continuation provides a natural (preferred) regularization of each factor.\\

{\em Remark 2.2:} The QFT with scale invariant correlation functions (2.2)
and (2.7) can be viewed as a (local chart) section of a M\"obius (i.e.
$SL(2,\R) \times SL(2,\R))$ invariant theory defined on conformally
compactified Minkowski space whose chiral projection is the circle $\SI^1$.
To display this fact we set
$$
   x = x(\xi) = 2\,\mbox{tg}\frac{{\scriptstyle \xi}}{^{\scriptstyle 2}} 
          \Longleftrightarrow
          e^{-i\xi} = \frac{1-\frac{i}{2}x}{1+\frac{i}{2}x}\  
          (= 1 - ix - \frac{x^2}{2} + \dots) \eqno{\mbox{(2.11a)}}
$$
for real $x$ and
$$
   \eps - ix = 2\,\mbox{th}\frac{{\scriptstyle \tilde{\eps} - i\xi}}
                            {^{\scriptstyle 2}}
          \Longleftrightarrow e^{\tilde{\eps} -i\xi}
          = \frac{1 + \frac{\eps - ix}{2}}{1 - \frac{\eps - ix}{2}}\,;
          \eqno{\mbox{(2.11b)}}
$$
the fields should also change under the general reparametrization law
\addtocounter{equation}{2}
\begin{eqnarray}
   j, \psi \rightarrow J(\xi) = 2\pi x'(\xi)\,j(x(\xi)),\ 
x'(\xi) = \left(\cos\frac{{\scriptstyle \xi}}{^{\scriptstyle 2}}\right)^{-2},
                                                         \nonumber \\
   \Psi(\xi,g) = 
   \sqrt{2\pi}\left(\cos\frac{{\scriptstyle \xi}}
                {^{\scriptstyle 2}}\right)^{-g^2}
   \psi(2\,\mbox{tg}\frac{{\scriptstyle \xi}}{^{\scriptstyle 2}}, g)\,.
\end{eqnarray}
The compact picture correlation functions are then obtained from (2.2), (2.7)
by the substitution
\begin{equation}
   (\eps - i\xx)^\lambda \rightarrow 
    \left(2\,\mbox{sh}\frac{{\scriptstyle \tilde{\eps} -i\xi_{12}}}
               {^{\scriptstyle 2}}\right)^\lambda\,.
\end{equation}
We note the change of the regularization parameter $(\eps \rightarrow
\tilde{\eps})$. It corresponds to a change of the energy operator in the
passage from the non-compact to the compact picture. While non-compact
energy is a Minkowskian time $(x^0)$ translation generator (which is an
isometry of the line element $dx$ since $x = x^1 - x^0$) the compact
picture energy (the Virasoro generator $L_0$) shifts the time parameter
on the circle and is hence an isometry of $d\xi$. {\em The ``preferred''
regularization is thus linked to the definition of energy.}\\

The (``observable'') current $J$ is periodic in $\xi$,
\begin{equation}
   J(\xi + 2\pi) = J(\xi) = \sum_{n \in \ZZ}\,J_n\,e^{in\xi}
\end{equation}
and can hence be viewed as a field on the circle. By contrast, the charged
field $\Psi(\xi,g)$, for non-integer dimension $\frac{1}{2}g^2$, only
obeys a twisted periodicity condition
\begin{equation}
   \Psi(\xi + 2\pi,g) = e^{-i\pi g^2}\,\Psi(\xi,g)\,e^{-2\pi igJ_0}
\end{equation}
where $J_0$ is the ``charge operator'' (the zero mode in the expansion
(2.14)). It should be interpreted as a section of a line bundle on $\SI^1$.
The exchange relations in the resulting {\em compact picture} assume the
form
\begin{equation}
   \Psi(\xi_2 + i\eps,g_2)\Psi(\xi_1,g_1)|\Phi\rangle =
    U^c_\eps(\xi_{12}, g_1g_2)\,
       \Psi(\xi_1 + i\eps,g_1)\Psi(\xi_2,g_2)|\Phi\rangle
\end{equation}
where we have dropped the tilde sign over $\eps$ and set
\begin{equation}
   U^c_\eps(\xi,\lambda) = 
     \left( \frac{e^{i\xi+\eps} - 1}{1 - e^{i\xi-\eps}}\right)^\lambda\,
         e^{-\eps\lambda}
     = e^{-2\pi i\lambda}\,U^c_\eps(\xi + 2\pi,\lambda)\;.
\end{equation}
%
\section{Consistent computations involving the exchange relations and the
equations of motion}
%
A closer look at the exchange relations (2.9) reveals an irritating problem
when we remove the regularization, let $\eps \rightarrow 0$ and go over to
a distributional interpretation. Even in the simplest case of (2.9) with 
$g_1 = -g_2$ applied to the 2-point function (2.8) in the variable 
$x = \xx = x_1 - x_2$, a naive passage to the 
limit produces the formula
\begin{equation}
    (0 + ix)^{-g^2} = U(x,-g^2)(0-ix)^{-g^2}\,.
\end{equation}
But the right hand side in (3.1) is ill defined as a distribution, being a 
product of a distribution with the discontinuous function
\begin{equation}
   U(x,\lambda) = \left(\frac{0+ix}{0-ix}\right)^\lambda
    = e^{i\pi\lambda}\theta(x) + e^{-i\pi\lambda}\theta(-x)
\end{equation}
where $\theta$ denotes the Heaviside function. We shall demonstrate below
how Eq. (3.1) is interpreted and handled correctly in the setting of
Colombeau algebras of generalized functions. First, however, we wish
to make the point that a careless interpretation of (3.1) within classical
distribution theory together with formal computations may lead to 
contradictions.\\

Indeed, differentiating (3.1) with respect to $x$ and applying the Leibniz
rule to the product on the right hand side we obtain
\begin{eqnarray*}
   i\frac{d}{dx} (0+ix)^{-g^2} = g^2 (0+ix)^{-g^2-1} \\
     = \left(i\frac{d}{dx} U(x,-g^2)\right) (0-ix)^{-g^2}
                    - g^2 U(x,-g^2)(0-ix)^{-g^2-1}\;.
\end{eqnarray*}
Multiplying this by $(0-ix)^{g^2}$ and rearranging terms we arrive at
$$ 
   i\frac{d}{dx} U(x,-g^2) 
    = g^2\left(\frac{1}{0+ix} + \frac{1}{0-ix}\right) U(x,-g^2)
                                          \eqno{\mbox{(3.3a)}}
$$
$$
   = 2\pi g^2\delta(x)U(x,-g^2)         \eqno{\mbox{(3.3b)}}
$$
which contains again a non-defined product of distributions. On the other
hand, if one uses the form (3.2) of $U(x,-g^2)$, one arrives at
\addtocounter{equation}{1}
\begin{equation}
   i\frac{d}{dx} U(x,-g^2) = 2\sin(\pi g^2)\delta(x)\;.
\end{equation}
Equating (3.3b) with (3.4) we see that, formally,
\begin{equation}
   2\pi g^2\delta(x)U(x,-g^2) = 2\sin(\pi g^2)\delta(x)\;.
\end{equation}
Taking the complex conjugate of Eq. (3.5) and multiplying by
$U(x,-g^2)$ we would arrive at
\begin{equation}
   2\pi g^2\delta(x) = 2\sin(\pi g^2)\delta(x)U(x,-g^2)
\end{equation}
which clearly contradicts (3.5). A similar problem arises if we use the 
compact picture (respectively Eq. (2.16) instead of (2.9)).\\

We shall now re-examine the situation from the view-point of algebras
of generalized functions and show how to avoid the contradiction.\\

Our first task is to define $U(x,\lambda)$ as an element of the 
Colombeau algebra $\cG(\R)$. This
will be simply done by taking the regularization suggested in (2.10) 
as a representing sequence, that is, $U(x,\lambda)$ will be the class of
$(U_\eps(x,\lambda))_{\eps>0}$ with $U_\eps(x,\lambda)$ defined by (2.10a) 
and $\sigma_\eps(x)$ defined by (2.10b). Using the principal branch of the
logarithm, it is easy to see that $\sigma_\eps(x)$ is a smooth, bounded
function of the real argument $x$ and satisfies the required bounds (A.1)
in terms of $\eps > 0$ (see the Appendix). Thus the same is true of
$U_\eps(x,\lambda)$ and so the respective classes correctly define the 
elements $U(x,\lambda)$ and $\sigma(x)$ in the algebra $\cG(\R)$. In a
similar way, the distributions $(0+ix)^{-g^2}$ and $(0-ix)^{-g^2}$
are interpreted as elements of $\cG(\R)$, and the equation (3.1) holds strictly
as an equality in the algebra $\cG(\R)$. Further, we have the identity
\begin{equation}
   U(x,-g^2) = e^{{\textstyle -ig^2\sigma(x)}}
\end{equation}
which can be differentiated using the chain rule, yielding
\begin{equation}
  i\frac{d}{dx} U(x,-g^2) = g^2\sigma'(x)U(x,-g^2)\;.
\end{equation}
We note in passing that (3.8) can be derived from Eq. (3.1) in $\cG(\R)$
as well. To see what it says in terms of distributions, we go over to the
association (see Appendix). The limiting behavior of $\sigma_\eps(x)$ expressed in
(2.10b) just says that
\begin{equation}
    \sigma(x) \approx \pi\theta(x) - \pi\theta(-x)\;.
\end{equation}
Association relations may be differentiated, so that
\begin{equation}
   \sigma'(x) \approx 2\pi\delta(x)\;.
\end{equation}
Now we can state the correct interpretation of Eq. (3.3) in the
algebra $\cG(\R)$: The first equality in (3.3a) is just (3.8),
whereas the second equality (3.3b) simply does not hold, because
$\sigma'(x) \neq 2\pi\delta(x)$ in $\cG(\R)$.\\ 

What concerns the remaining equations (3.2) and (3.4) - (3.6) we have 
from the identity (3.7):
\begin{equation}
   U(x,-g^2) = e^{{\textstyle -ig^2\sigma(x)}}
          \approx e^{{\textstyle -i\pi g^2}}\theta(x) 
            + e^{\textstyle i\pi g^2}\theta(-x)
\end{equation}
which replaces (3.2), and by differentiation
\begin{equation}
   i\frac{d}{dx} U(x,-g^2) \approx 2 \sin(\pi g^2)\delta(x)
\end{equation}
which replaces (3.4). Combining (3.11) with (3.12) we get
\begin{equation}
   g^2\sigma'(x)U(x,-g^2) \approx 2\sin(\pi g^2)\delta(x)
\end{equation}
which corresponds to (3.5). We know that an association relation
cannot be multiplied in general. Therefore, the manipulations leading
to (3.6) would be erronous in (3.13); thus the contradiction (3.6) 
cannot be derived in the setting of Colombeau algebras of generalized
functions.\\

Similarly, Eq. (2.8) and the anyonic exchange relations (2.9) are all
valid in the algebra $\cG(\R)$ (with $\xx = x_1 - x_2$), and (2.16), (2.17)
are valid in the sense of generalized fractional order densities on
the M\"obius strip. In addition,
individual terms have an interpretation by means of an associated 
distribution. Inconsistencies can only arise when the association relations
are unjustifiably combined with the differential-algebraic relations
expressed by the equations.\\

As a second application of the Colombeau approach, we shall now
specify the issues raised in Remark 2.1 and also comment on the
delta function terms seemingly appearing in Eq. (3.1) when $g^2$ is
an integer. We consider the one-dimensional distributions
$(0\pm ix)^\lambda$ arising from Wightman distributions in the
coordinate $x = x_1 - x_2$. As noted in Remark 2.1, the formulae
\begin{equation}
   (0+ix)^\lambda (0+ix)^\mu = (0+ix)^{\lambda+\mu},
   (0-ix)^\lambda (0-ix)^\mu = (0-ix)^{\lambda+\mu}
\end{equation}
have a meaning within Schwartz distribution theory, for instance by
Fourier transform and convolution. We have interpreted these distributions
as elements of the Colombeau algebra $\cG(\R)$ by means of their 
distinguished analytic regularization $(\eps \pm ix)^\lambda$. Then
formula (3.14) becomes
\[
   (\eps+ix)^\lambda (\eps+ix)^\mu = (\eps+ix)^{\lambda+\mu}
                    \approx (0+ix)^{\lambda + \mu}
\]
and similarly for the minus sign. Here the first equality holds strictly in
$\cG(\R)$, while the association reflects the fact that the product in
(3.14) has a distributional meaning. The situation is different when two
Wightman distributions with oppositely ordered arguments are multiplied.
The product has of course a meaning in the algebra $\cG(\R)$, but will not
admit an associated distribution. For example,
\[
   \frac{1}{\eps+ix}\cdot\frac{1}{\eps-ix} = \frac{1}{\eps^2+x^2}
\]
holds as an identity in $\cG(\R)$, but defines a non-distribution generalized
function with diverging representative, as is seen from the fact that
\[
   \eps\frac{1}{\eps^2+x^2} \approx \pi\delta(x)\;.
\]

The appearance of the delta function terms in the integer-valued case can 
easily be explained in this context. Indeed, we can write
\begin{eqnarray}
  (\eps+ix)^{-k} \approx (-i)^k x^{-k} + 
       \frac{i^{k-1}\pi}{(k-1)!}\delta^{(k-1)}(x) = (0 + ix)^{-k}\,, 
                                                       \nonumber \\
  (\eps-ix)^{-k} \approx i^k x^{-k} + 
       \frac{(-i)^{k-1}\pi}{(k-1)!}\delta^{(k-1)}(x) = (0 - ix)^{-k}\;.
\end{eqnarray}
The distribution $x^{-k}$ is understood in the sense of Hadamard's finite
part, or equivalently as
\[
   x^{-k} = \frac{(-1)^{k-1}}{(k-1)!}\;\frac{d^k}{dx^k}\ln|x|\;.
\]
The exchange relations
\[
    (\eps + ix)^{-k} = U_\eps(x,-k)(\eps - ix)^{-k}
\]
thus imply that
\begin{equation}
   U_\eps(x,-k)(\eps - ix)^{-k} \approx (0+ix)^{-k}\;.
\end{equation}
On the other hand, $U_\eps(x,-k) \approx (-1)^k$ and so one might be tempted
to write $U_\eps(x,-k)(\eps-ix)^{-k} \approx (-1)^k(0-ix)^{-k}$. This,
however, is an illegitimate multiplication in an association relation. It
holds only in the open sets $x > 0$ or $x < 0$. The correct extension to
$x = 0$ is obtained from combining Eqs. (3.15) and (3.16) and just says that
\[
   (-1)^k(0-ix)^{-k} = (0+ix)^{-k} - 
             \frac{2i^{k-1}\pi}{(k-1)!} \delta^{(k-1)}(x)\;.
\]  
{\bf Concluding remarks} 
\vspace{3mm}\\
The concept of generalized functions in the 
sense of Colombeau distinguishes between sequences of smooth functions
which have the same limit in the framework of distribution theory. Thus,
it is adapted to exploit the presence of distinguished (preferred)
regularizations. Such a distinguished regularization does exist in QFT
where the basic requirement of energy positivity guarantees that
Wightman distributions can be viewed as boundary values of analytic
functions holomorphic in a tube domain.\\

The Colombeau algebras offer a framework in which the generalized functions
arising in QFT can be given a meaning. In this setting, the usual rules
of analysis (differentiation, multiplication, Leibniz rule, chain rule)
are all valid and can be applied without restriction. The important
fundamental equations continue to hold. Care is only needed when
re-interpreting the results in terms of classical distribution theory.
Via the concept of association, individual generalized functions correspond
to the distributions known from QFT. However, inserting these associated
distributions back into the equations may lead to ill-defined products
(on the level of distribution theory) and is thus not allowed. In short,
differential-algebraic computations have to be done on the level of
algebras of generalized functions, the interpretation of the
resulting solutions can be done on the level of classical distribution
theory. Further, the rules accompanying the concept of association may 
guide us in deciding what equations remain valid as distributional
equations and which ones hold only in the sense of algebras of generalized
functions.\\

To conclude, the Colombeau algebras of generalized functions appear to be
more flexible and better adapted for applications in QFT than the currently
used Schwartz distributions. Our analysis confirms the common sense rule 
that the basic mathematical concepts used in such a rich and unsettled
domain of modern physics as QFT should not be viewed as rigidly fixed once
for all, but should reflect significant new developments in both mathematics
and physics.
\vspace{7mm}\\
{\bf Acknowledgements}
\vspace{3mm}\\
Two of the authors, M. O. and I. T., would like to acknowledge the
hospitality of the Erwin Schr\"odinger International Institute for Mathematical Physics
in Vienna where this work was conceived and a major part of it was done.
\vspace{1mm}
\pagebreak\\
%
{\bf \large Appendix: Algebras of generalized functions}
%
\vspace{5mm}\\
We sum up here some basic facts about Colombeau algebras of generalized
functions \cite{Colo} used in the text.\\

Our aim is to embed the space $\cD'$ of Schwartz distributions into a
larger (commutative and associative) algebra $\cG$ in which differentiation
and nonlinear superposition can be performed.\\

A well known result of Schwartz \cite{Sch1} says that any embedding of
$\cD'$ into an associative, commutative differential algebra that
preserves derivatives must necessarily change the product of continuous
functions. The Colombeau construction is optimal in as much as it
preserves the product of smooth functions.\\

We shall outline the construction of $\cG$ on the real line first.
Generalizations to higher dimensions and to generalized sections of
vector bundles will be briefly indicated at the end of the Appendix.
For further details and for the concept of an operator valued generalized
function we refer to the monographs \cite{Colo,Ober}.\\

Every distribution $w \in \cD'(\R)$ can be approximated by a sequence of
smooth regularizations $(w_\eps)_{\eps > 0}$. Identifying the regularizing
sequences with the same limit, we see that $\cD'(\R)$ can be described as
the factor space $\cV(\R)/\cV_0(\R)$, where $\cV(\R)$ denotes the space of
distributionally convergent sequences of smooth functions and $\cV_0(\R)$
the zero sequences. The linearity inherent in $\cD'(\R)$ is reflected by
the fact that $\cV(\R)$ is not an algebra (for example, the square of
a real valued sequence converging to the Dirac delta function never belongs
to $\cV(\R)$). What is more, $\cV_0(\R)$ can not be contained in a proper
ideal in whatever algebra. This is clear from the observation that the
sequences $(\sin\frac{x}{\eps})_{\eps > 0}$ and 
$(\cos\frac{x}{\eps})_{\eps > 0}$ converge to zero in the sense of 
distributions, i.~e. belong to $\cV_0(\R)$, but satisfy 
$(\sin\frac{x}{\eps})^2 + (\cos\frac{x}{\eps})^2 = 1$. Thus our strategy
will be to enlarge $\cV(\R)$ so that we obtain an algebra and to make 
$\cV_0(\R)$ smaller to get an ideal.\\

We introduce the algebra $\cE(\R)$ of sequences $(u_\eps)_{\eps > 0}$ of
smooth functions with the property of {\em moderate growth} in $\eps$:
For every compact set $K \subset \R$ and every integer $k \geq 0$ there
is $N \geq 0$ such that
$$
  \left|\frac{d^k}{dx^k}u_\eps(x)\right| = \cO(\eps^{-N}) \eqno{\mbox{(A.1)}}
$$
as $\eps \rightarrow 0$, uniformly for $x \in K$. Further, $\cN(\R)$ will
denote the sequences $(u_\eps)_{\eps > 0}$ enjoying the property of 
{\em rapid decrease} in $\eps$: For every compact set $K \subset \R$,
every integer $k \geq 0$ and every $M > 0$,
$$
  \left|\frac{d^k}{dx^k}u_\eps(x)\right| = \cO(\eps^M)   \eqno{\mbox{(A.2)}}
$$
as $\eps \rightarrow 0$, uniformly for $x \in K$. With componentwise 
operations, $\cE(\R)$ is a differential algebra and $\cN(\R)$ a differential
ideal. Thus the factor algebra
$$
   \cG(\R) = \cE(\R)/\cN(\R)   \eqno{\mbox{(A.3)}}
$$
is a differential algebra. We note that sequences $(w_\eps)_{\eps>0}$
obtained by convolution of a distribution $w$ with a smoothing kernel
(also called a {\em mollifier})
from $\cE(\R)$ belong to $\cE(\R)$ as well. Further, the ideal
$\cN(\R)$ is contained in $\cV_0(\R)$. Thus the assignement
$$
   \cD'(\R) \rightarrow \cG(\R)\,:\, 
      w \rightarrow \ \mbox{class of}\ (w_\eps)_{\eps>0}  \eqno{\mbox{(A.4)}}
$$
defines an embedding of the space of distributions into the algebra
$\cG(\R)$ which preserves differentiation. Further, the class of smoothing
kernels can be suitably chosen so that (A.4) also preserves the product of
smooth functions. The superposition $F(u)$ is a well-defined element of
$\cG(\R)$, provided $F$ is smooth and polynomially bounded (then $u$ may
be an arbitrary element of $\cG(\R)$) or else $u$ is a bounded generalized
function (with respect to $\eps$) and $F$ is arbitrary. Thus the algebra
$\cG(\R)$ provides a framework in which nonlinear operations with
distributions as well as differentiation can be performed.\\

In order to relate elements of $\cG(\R)$ to distributions, whenever
possible, we introduce the notion of association. An element $u =$ class
of $(u_\eps)_{\eps>0}$ of $\cG(\R)$ is said to {\em admit an associated
distribution} $w$, notation $u \approx w$, provided
\[
   \lim_{\eps \to 0}\,u_\eps = w
\]
in the distributional sense.\\

As a simple application, we show how this concept allows to avoid the
Schwartz impossibility result. Consider the three distributions
$1/x = (\ln|x|)'$, $x$, $\delta(x)$. We view them as elements of
$\cG(\R)$ via the embedding (A.4). By associativity, it holds that
$$
   \big( \frac{1}{x} \cdot x \big) \cdot \delta(x)
      = \frac{1}{x} \cdot \big( x \cdot \delta(x) \big)\;.  
                                                \eqno{\mbox{(A.5)}}
$$
Within classical distribution theory, $\frac{1}{x}\cdot x = 1$ and
$x \cdot \delta(x) = 0$, seemingly leading to the contradiction
$\delta(x) = 0$ by (A.5). The contradiction does not arise in $\cG(\R)$,
because there $\frac{1}{x}\cdot x \neq 1$ and $x \cdot \delta(x) \neq 0$.
The distributional relations continue to hold in the sense of association,
though: $\frac{1}{x}\cdot x \approx 1, x \cdot \delta(x) \approx 0$.
The crucial point is that we are not allowed to perform the algebraic
operations arising in Eq. (A.5) in the association relations above. 
The association relation is compatible with differentiation, but not with
multiplication. As a general rule, algebraic operations should take place 
in $\cG$, while a distributional interpretation of individual terms 
(separately) is supplied by association.\\

We now turn to Colombeau algebras on manifolds. First, if $\Omega$ is an
open subset of $\R^n$, the definitions (A.1) - (A.3) extend in a
straightforward manner to give the algebra $\cG(\Omega)$. The embedding
(A.4) works on compactly supported distributions and can be extended
uniquely to an embedding of $\cD'(\Omega)$ into $\cG(\Omega)$ by
localization. Next, let $(E,X,p)$ be a vector bundle with base space
$X$ a smooth $n$-dimensional manifold, projection $p: E \to X$ and fiber
$\R^m$. Vector bundle charts will be written in the form
\begin{eqnarray*}
  \kappa : p^{-1}(V) & \rightarrow & \kappa_0(V) \times \R^m\,,\\
              z & \rightarrow & (\kappa_0(p(z)), \vec{\kappa}(z))
\end{eqnarray*}
where $\kappa_0$ is a chart of the manifold $X$ with domain $V$. Given a
sequence of smooth sections $u_\eps : X \to E$ and a vector bundle chart
$\kappa$, we can consider the maps
$$
   \vec{\kappa} \circ u_\eps \circ \kappa^{-1}_0\,:\,
           \kappa_0(V) \rightarrow \R^m\;.          \eqno{\mbox{(A.6)}}
$$
Let $\cE(X;E)$ be the space of sequences $(u_\eps)_{\eps>0}$ such that
each composition (A.6) satisfies a bound of type (A.1). Similarly, 
$\cN(X;E)$ is defined by requiring the bounds (A.2) in (A.6). The space of
generalized sections is defined by
\[
   \cG(X;E) = \cE(X;E)/\cN(X;E)\,.
\]
If $E$ is the trivial bundle $E = X \times \R$ we obtain the
Colombeau algebra $\cG(X)$ of generalized functions on the manifold $X$. 
The embedding of the space of distributions $\cD'(X)$ into $\cG(X)$ is
more delicate and requires the choice of an atlas and a subordinate
partition of unity.\\

In the compact picture, for example, the current in (2.12) defines a 
generalized 1-form on the circle, while the charged fields
$\Psi(\xi,g)$ of non-integer dimension $\frac{1}{2}g^2$ can be
interpreted as fractional order densities on the M\"obius strip.


\begin{thebibliography}{10}

\bibitem{Bog1} {\sc N.N. Bogolubov, A.A. Logunov, A.I. Oksak, 
I.T. Todorov:} General Principles of Quantum Field Theory.
Kluwer Acad. Publ., Dordrecht 1990.

\bibitem{Bog2} {\sc N.N. Bogolubov, D.V. Shirkov:} Introduction
to the Theory of Quantized Fields. Wiley-Interscience, New York 1959.

\bibitem{Buch} {\sc D. Buchholz, G. Mack, I.T. Todorov:} The current
algebra on the circle as a germ of local field theories. 
Nucl. Phys. B (Proc. Suppl), {\bf 5B} (1988), 20 - 56.
 
\bibitem{Coll} {\sc J. Collins:} Renormalization. Cambridge Univ. Press,
Cambridge 1984.

\bibitem{Colo} {\sc J.F. Colombeau:} New Generalized Functions and 
Multiplication of Distributions. North-Holland,
Amsterdam 1985.
 
\bibitem{Fran} {\sc P. Di Francesco, P. Mathieu, D. Senechal:} Conformal
Field Theory. Springer, Berlin 1997.

\bibitem{Haag} {\sc R. Haag:} Local Quantum Physics. Springer, Berlin 1992.

\bibitem{Ilie} {\sc N. Ilieva, W. Thirring:} The Thirring model 40 years
later. Vienna preprint UWThPh-1998-37.

\bibitem{Ober} {\sc  M. Oberguggenberger:} Multiplication of Distributions and
Application to Partial Differential Equations. 
Longman, Harlow 1992.
 
\bibitem{Rosi} {\sc E.E. Rosinger:} Nonlinear Partial Differential Equations:
An Algebraic View of Generalized Functions. North-Holland, Amsterdam 1990.

\bibitem{Sch1} {\sc L. Schwartz:} Sur l'impossibilit\'e de la multiplication
des distributions. C. R. Acad. Sci. Paris {\bf 239} (1954), 847 - 848.

\bibitem{Sch2} {\sc L. Schwartz:} Th\'eorie des distributions: Nouvelle ed.,
Hermann, Paris 1966.

\bibitem{Stre} {\sc R.F. Streater, A.S. Wightman:} PTC, Spin and Statistics,
and All That. 2nd ed., Benjamin/Cummings, Reading, MA 1978.

\bibitem{Thir} {\sc W. Thirring:} A soluble relativistic field theory.
Ann. Phys. {\bf 3} (1958), 91 - 112.

\end{thebibliography}
\end{document}